\documentstyle[12pt]{article}

\newcounter{subseqn}
\newcounter{saveeqn}

\renewcommand{\thesection}%
 {\Roman{section}.\setcounter{equation}{0}}
\renewcommand{\thesubsection}%
 {\setcounter{subseqn}{\value{equation}}%
\thesection\Roman{subsection}%
\setcounter{equation}{\value{subseqn}}}
\renewcommand{\theequation}%
 {\mbox{\arabic{section}.\arabic{equation}}}

\newcommand{\alpheqn}%
 {\setcounter{saveeqn}{\value{equation}}%
\stepcounter{saveeqn}\setcounter{equation}{0}%
\renewcommand{\theequation}%
 {\mbox{\arabic{section}.\arabic{saveeqn}\alph{equation}}}}

\begin{document}

\title{$\delta$-Expansion at Finite Temperature}

\author{Rudnei O. Ramos\thanks{E-mail: rudnei@vmesa.uerj.br} \\
{\it Universidade do Estado do Rio de Janeiro, }\\
{\it Instituto de F\'{\i}sica - Departamento de F\'{\i}sica Te\'orica,}\\
{\it 20550-013 Rio de Janeiro, RJ, Brazil}}

\date{IF-UERJ 13/96 , February 1996}


\maketitle
\thispagestyle{empty}

\begin{abstract}

\baselineskip 24pt

We apply the $\delta$-expansion perturbation scheme to the $\lambda \phi^{4}$
self-interacting scalar field theory in 3+1 D at finite temperature. In the
$\delta$-expansion the interaction term is written as $\lambda (\phi^{2})^{
1 + \delta}$ and $\delta$ is considered as the perturbation parameter. We
compute within this perturbative approach the renormalized mass at finite
temperature at a finite order in $\delta$. The results are
compared with the usual loop-expansion at finite temperature.


\end{abstract}

\newpage
\setcounter{page}{1}

\section{Introduction}

The study of field theories at finite temperatures has long been an
important issue in high energy physics (for a general review see
\cite{kapusta}).
However, in many situations, when working with field theories at finite
temperatures, usual perturbations schemes break down due to the appearance
of infrared divergences (for example, close to critical temperatures, in
field theories with symmetry breaking and for massless field theories, like
QCD, or for small values of "effective masses"). In those situations, we
must usually perform a resummation procedure to take in to account
relevant contributions in the infrared region (see for example \cite{dolan})
or make use of nonperturbative approaches for studying the theory in the
infrared region, as a renormalization group study or by the
$\epsilon$-expansion technique, for example.

Recently a new perturbation scheme in field theory was proposed, known as the
\break $\delta$-expansion \cite{bender,yotsu}. In this novel
perturbation
scheme, instead of using Lagrangian parameters for the expansion, like an
expansion in the interaction coupling constant $\lambda$ in the $\lambda
\phi^{4}$ theory (regarding $\lambda$ as a weak-coupling constant) or the
usual loop-expansion (in powers of $\hbar$) the $\delta$-expansion makes use
of an artificial parameter ($ \delta$).

In the usual $\lambda \phi^{4}$ theory in 3+1D, the interaction term is
rewritten as $\lambda M^{4} (M^{-2} \phi^{2})^{1 + \delta}$, where $M$ is an
arbitrary mass parameter introduced to make the coupling
constant $\lambda$ dimensionless. $\delta$ is regarded as a small positive
parameter that can be used as a perturbative parameter in the theory, for
example when Green's functions are computed.

If one expands the interaction term in powers of $\delta$ we get
$\lambda \phi^{4} \to \lambda M^{2} \phi^{2} + \lambda
M^{2} \phi^{2} \sum_{n=1}^{\infty} \frac{\delta^{n}}{n !} \left[ \ln (M^{-2}
\phi^{2}) \right]^{n}$. Therefore, the $\delta$-expansion generates a mass
term which can not only make the behavior of the theory in the infrared
region better, but also introduces nonperturbative effects in the coupling
$\lambda$ once $M$ is fixed according to an appropriate procedure, as
described below.

We recover the original interaction term for $\delta = 1$ and the dependence on
the arbitrary mass parameter $M$ goes away. However, in this paper, we
will be interested exactly in what happens when we keep the $\delta$-expansion
up to a finite order in $\delta$ and when the results carry a dependence on
$M$. We will be particularly interested in computing the renormalized mass
$m_{R}$, at finite temperature, in the $\lambda \phi^{4}$ scalar model
at some finite order in $\delta$. Since in this case $m_{R}$ is dependent on
$M$, we must choose an optimization scheme to fix the value of $M$. Here
we choose the Principle of Minimal Sensitivity (PMS) \cite{stevenson} ,
where the
quantities we are interested in are required to be stationary with respect
to $M$.
We are going to show that, in the evaluation of the renormalized mass
at finite temperature, by fixing the mass parameter $M$ through this
variational method, we obtain a gap equation for the effective mass at finite
temperature without having to use the usual resummation of diagrams
(see, for instance, \cite{dolan}).

The paper in organized in the following way: In Section II, we give a brief
review of the $\delta$-expansion technique and show how to compute the
vertex Green's functions at finite temperature. In Section III, we 
demonstrate the
procedure by computing the effective mass at finite temperature and obtain
the gap equation. In Section IV, we have our conclusions and comments on
further applications of the method.

\section{The $\delta$-Expansion Approach: Computing Green's Functions at
Finite Temperature}

We begin by giving a short review of the $\delta$-expansion approach. The
$\lambda \phi^{4}$
Lagrangian density for a scalar field $\phi$, in 3+1D, given by

\begin{equation}
{\cal L} = \frac{1}{2} (\partial_{\mu} \phi)^{2} - \frac{\mu^{2}}{2} \phi^{2}
- \frac{\lambda}{4 } \phi^{4}
\label{e1}
\end{equation}

\noindent
is rewritten as

\begin{equation}
{\cal L} = \frac{1}{2} (\partial_{\mu} \phi)^{2} - \frac{\mu^{2}}{2} \phi^{2}
- \frac{\lambda}{4 } M^{4} (M^{-2} \phi^{2})^{1 + \delta} \: .
\label{e2}
\end{equation}
 
\noindent
Expanding (\ref{e2}) in powers of $\delta$ we get

\begin{equation}
{\cal L} = \frac{1}{2} (\partial_{\mu} \phi)^{2} - \frac{1}{2} ( \mu^{2} +
\frac{\lambda}{4 } 2 M^{2} ) \phi^{2} - \frac{\lambda}{4 } M^{2}
\phi^{2} \sum_{n=1}^{\infty} \frac{\delta^{n}}{n !} \left[ \ln (M^{-2}
\phi^{2}) \right]^{n} \: .
\label{e3}
\end{equation}

\noindent
If one uses that

\begin{equation}
\sum_{n=0}^{\infty} \frac{\delta^{n}}{n !} \left[ \ln (M^{-2} \phi^{2})
\right]^{n} = \sum_{n=0}^{\infty} \frac{\delta^{n}}{n !}
\frac{ d^{n}}{dk^{n}} (M^{-2} \phi^{2})^{k} |_{k=0} =
e^{\delta \partial_{k}} (M^{-2} \phi^{2})^{k} |_{k=0} \:,
\label{e4}
\end{equation}

\noindent
then (\ref{e3}) can be written as \cite{yotsu}

\begin{equation}
{\cal L} = \frac{1}{2} (\partial_{\mu} \phi)^{2} - \frac{1}{2} ( \mu^{2} +
\frac{\lambda}{4 } 2 M^{2} ) \phi^{2} - D_{k} \phi^{2k+2} |_{k=0} \:,
\label{e5}
\end{equation}

\noindent
where, from the relation (\ref{e4}), $D_{k}$ is a derivative operator given by

\begin{equation}
D_{k} = \frac{\lambda}{4 } M^{2} \left( e^{\delta \partial_{k}} - 1
\right) \left( M^{-2} \right)^{k} \: .
\label{e6}
\end{equation}

In ref. \cite{yotsu} it was shown that the $n$-point Green's function $G^{(n)}
( x_{1}, x_{2}, \ldots , x_{n})$ can be written as\footnote{In ref.
\cite{bender} the Green's functions are defined differently but
the final results are completely analogous.}

\begin{eqnarray}
\lefteqn{G^{(n)} ( x_{1}, x_{2}, \ldots , x_{n})  =
\prod_{p=0}^{\infty} \frac{1}{p !} \int d^{4}y_{1} d^{4} y_{2} \ldots
d^{4} y_{p} \langle 0 | T \phi(x_{1}) \phi(x_{2}) \ldots \phi(x_{n})
\times} \nonumber \\
& & D_{k_{1}} D_{k_{2}} \ldots D_{k_{p}} \left[ \phi^{2}(y_{1})
\right]^{k_{1}+1} \left[ \phi^{2}(y_{2})\right]^{k_{2}+1} \ldots
\left[ \phi^{2}(y_{p})\right]^{k_{p}+1} | 0 \rangle_{c} |_{k=0} \: ,
\label{e7}
\end{eqnarray}

\noindent
which can be computed, as shown in ref. \cite{yotsu}, by first considering
the $k_{i}$'s as integers with the same value such that we can draw all
diagrams coming from (\ref{e7}). From (\ref{e6}), if the $k$'s 
are integers then $D_{k}$ can
be regarded as small and (\ref{e7}) can be computed by ordinary diagrammatic
perturbation. At the end, considering the k's as continuous with $k_{i} \neq
k_{j}, \: i \neq j$, we apply the derivative operators $D_{k_{i}}$ and finally
we make all $k$'s igual to zero.

Once we know how to compute the Green's functions, we can obtain the
renormalized mass $m_{R}$, the renormalized coupling constant $\lambda_{R}$ and
the wave-function renormalization constant $Z$ from the usual definitions:

\begin{equation}
m_{R}^{2} = Z \left[ G_{c}^{(2)} (p^{2}) \right]^{-1} |_{p^{2}=0} \: ,
\label{e8}
\end{equation}

\begin{equation}
\lambda_{R} = - Z^{2} G_{c}^{(4)} (0,0,0,0)  \: ,
\label{e9}
\end{equation}

\noindent
and

\begin{equation}
Z^{-1} = 1 + \frac{d}{d p^{2}} \left[ G_{c}^{(2)} (p^{2}) \right]^{-1}
|_{p^{2}=0} \: ,
\label{e10}
\end{equation}

\noindent
where $G_{c}^{(2)} (p^{2})$ and $G_{c}^{(4)}(p_{1}, p_{2}, p_{3}, p_{4})$
are the connected two-point and four-point Euclidean Green's functions, in
momentum space, respectively.


At lowest order ($\lambda$), the two-point Green's
function $G^{(2)}$ is given by an one-vertex diagram as \cite{yotsu}

\begin{equation}
G_{(1v)}^{(2)} = - D_{k_1} \frac{ (2k_1 + 2) !}{2^{k_1} k_1 !} \left[
I(m) \right]^{k_1} |_{k_1 = 0} \: ,
\label{G1}
\end{equation}

\noindent
where $D_{k_1}$ is given by (\ref{e6}) and $I(m)$ is the usual loop integral,
at $T \neq 0$, given by \cite{dolan}

\begin{equation}
I(m) = \frac{1}{\beta} \sum_{n=-\infty}^{n= + \infty} \int \frac{ d^{3} q}{
(2 \pi)^{3}} \frac{1}{\omega_{n}^{2} + q^{2} + m^{2}} \: ,
\label{Im1}
\end{equation}

\noindent
where, from (\ref{e3}), $m^{2} = \mu^{2} + \frac{\lambda}{4 } 2 M^{2}$,
$\beta =
T^{-1}$ is the inverse of the temperature and $\omega_{n} = \frac{2 \pi n}
{\beta}$.

Subtracting the zero temperature divergent contribution\footnote{
{}From the expressions we will obtain later, it is straightforward to
show that renormalization by the introduction of counterterms in the
Lagrangian
density is enough. Thus we are going to refer only to the finite ($T\neq 0$)
contributions. {}For a discussion of the renormalization up to order
$\delta^2$, at
$T=0$, see \cite{yotsu} and Bender and Jones in \cite{bender}.} of (\ref{Im1}),
one can write the
following expansion \cite{braden}
for $I(m)$ in powers of $m^{2} \beta^{2}$:

\begin{equation}
I(m) = \frac{T^{2}}{12} - \frac{m T}{4 \pi} - \frac{m^{2}}{8 \pi^{2}}
\left( \ln ( \frac{m}{4 \pi T}) + \gamma - \frac{1}{2} \right) +
\frac{m^{4}}{T^{2}} \frac{ \xi(3)}{2^{7} \pi^{4}} + {\cal O} \left(
m^4 \beta^4 \right) \: .
\label{Im2}
\end{equation}

A consistent evaluation of the quantities (\ref{e8})-(\ref{e10})
at an order
higher than $\delta$ must also include the evaluation of Green's
functions of an equivalent order in the number of vertices.

The two-vertex
Green's function, $G_{(2v)}^{(2)}$, from (\ref{e7}), would be given by
(including symmetry factors)

\begin{eqnarray}
G_{(2v)}^{(2)} &=& \frac{1}{2 !} \sum_{n = 0}^{2} \frac{2 !}{(2 - n)!
n !} D_{k_1} D_{k_2} \sum_{l=2}^{+\infty} \frac{(2 k_1 + 2)!}{
2^{k_1 + \frac{(n - l)}{2}} \left(k_1 + \frac{(n - l)}{2}\right)!}
\left[ I(m) \right]^{k_1 + \frac{(n - l)}{2}} \times
\nonumber \\
& \times & \frac{(2 k_2 + 2)!}{2^{k_2 + 1 - \frac{(n + l)}{2}}
\left( k_2 + 1 - \frac{(n + l)}{2} \right)!}
\left[ I(m) \right]^{k_2 + 1 - \frac{(n + l)}{2}}
\frac{J_{l}(\tilde{p})}{l !} |_{k_1,k_2 = 0} \:,
\label{G2}
\end{eqnarray}

\noindent
where, (\ref{G2}) must be evaluated subject to the following constraint
in the positive integer numbers $n$ and $l$: $n+l$ must be
even in order to have an even number of field lines leaving each vertex.
In (\ref{G2}), $J_l (\tilde{p})$ represents internal
propagators (between vertices) that at $T \neq 0$ are given by

\begin{equation}
J_l (\vec{\tilde{p}},\omega_n) = \prod_{i=1}^l \frac{1}{\beta} \sum_{n_i}
\int \frac{d^3 q_i}{(2 \pi)^3} \frac{\delta^3 (\vec{\tilde{p}} - \sum_{j=1}^l
\vec{q}_j) \delta_{n,\sum_j n_j}}{\omega_{n_i}^2 + \vec{q_i}^2 + m^2}
\label{Jl}
\end{equation}

\noindent
where $\omega_n = \frac{2 \pi n}{\beta}$, 
$\omega_{n_i} =\frac{2 \pi n_i}{\beta}$
are the Matsubara frequencies ($n$, $n_i$ $= 0,\pm 1, \pm 2, \ldots$).
$\tilde{p}$ (the external
momentum) is defined
by:

\begin{equation}
\tilde{p} = \left\{
\begin{array}{l}
0 \:, \: {\rm for}\: \;n=0 \:\; {\rm or }\: n=2; \\
p \:, \: {\rm for}\: \;n=1 \:.
\end{array}
\right.
\label{p1}
\end{equation}

Green's functions of higher vertices can be equivalently defined, using the
prescriptions given before.  However, for the purposes of this paper,
equations (\ref{G1}) and (\ref{G2}) will suffice for us.

Using Eq. (\ref{e6}) for the derivative operator $D_k$ in
(\ref{G1}) and  (\ref{G2}) and using that $e^{\delta \partial_k}
-1$ can be expanded as $\delta \frac{\partial}{\partial k} +
\frac{\delta^2}{2!} \frac{\partial^2}{\partial k^2} + \ldots$ , we
obtain for the two-point Green's function, up to two vertices, the
expression

\begin{eqnarray}
&G^{(2)}& = G_{(1v)}^{(2)} + G_{(2v)}^{(2)}  +
\ldots = D_{k_1} \tilde{G}_{(1v)}^{(2)}(k_1)|_{k_1 = 0} + D_{k_1} D_{k_2}
\tilde{G}_{(2v)}^{(2)}(k_1,k_2)|_{k_1,k_2 = 0}
 + \ldots = \nonumber \\
&=& \frac{\lambda}{4} M^2 \left( \delta \frac{\partial}{\partial k_1} +
\frac{\delta^2}{2!} \frac{\partial^2}{\partial k_1^2} +
 \ldots \right)
\left( M^{-2}\right)^{k_1} \tilde{G}_{(1v)}^{(2)}(k_1)|_{k_1 = 0} +
\nonumber \\
&+&
\frac{\lambda^2}{16} M^4 \left( \delta^2 \frac{\partial^2}{
\partial k_1 \partial k_2} + \ldots \right) \left( M^{-2}\right)^{k_1 + k_2}
\tilde{G}_{(2v)}^{(2)}(k_1,k_2)|_{k_1,k_2 = 0} +
\ldots \:,
\label{Gdelta3}
\end{eqnarray}

\noindent
where we have explicited the terms up to order $\delta^2$.
As an example of the evaluation procedure, we compute
below the effective mass at finite temperature and at order
$\delta^2$.

\section{Computing the Effective Mass at $T\neq 0$}

First, let us comment about expanding up to order $\delta^2$ in the limit of
high temperatures, $m \beta \ll 1$. When evaluating quantities like
effective masses up to order $\delta^2$,
for consistency we must also include contributions coming from the
two-vertex Green's function, since its leading term in $\delta$ is of order
$\delta^2$ (see Eq. (\ref{Gdelta3})). However, if we restrict our analyses
in the limit of high temperature and extending the results of \cite{yotsu}
for $G_{(2v)}^{(2)}$ up to order $\delta^2$ (evaluated in \cite{yotsu} at
zero temperature), a rather analogous evaluation at finite
temperature allows one to show that $G_{(2v)}^{(2)}$, when compared with
$G_{(1v)}^{(2)}$, contributes with
subleading corrections\footnote{An explicitly evaluation shows that
$G_{(2v)}^{(2)} \stackrel{m \beta \ll 1}{\longrightarrow}
\frac{\rm constant}{T^2}$, while, from Eq. (\ref{Gsimple}) and
(\ref{Im2}), $G_{(1v)}^{(2)} \stackrel{m \beta \ll 1}{\longrightarrow}
{\rm constant}\times (\ln T)^2$.} to the effective mass 
$m_{\rm eff}(T)$ at finite
temperature,
when we restrict the
evaluations in the high temperature limit $m\beta \ll 1$. For the same
reason above, since the wave function $Z^{-1}$, at order $\delta^2$, receives
contributions only
from $G_{(2v)}^{(2)}$, in the high temperature limit we
can write $Z^{-1} \stackrel{m\beta \ll 1}{\simeq} 1$.

We may, therefore, restrict just to (\ref{G1}), for $m \beta \ll 1$. Up to
second order in
$\delta$, we get the following expression for the
two-point Green's function given by (\ref{G1}):

\begin{eqnarray}
G^{(2)}_{(1v)} & =& - \frac{\lambda}{4 } 2 M^{2} \left\{
\delta \left[ \ln \left( \frac{I(m)}{2 M^{2}} \right) + 2 \psi(3) - \psi(1)
\right] + \frac{\delta^{2}}{2 !} \left[ \left[ \ln \left( \frac{I(m)}{2 M^{2}}
\right) + 2 \psi(3) - \psi(1)
\right]^{2}  \right. \right. \nonumber \\
&+& \left. \left. 4 \psi'(3) - \psi'(1) \right] \right\}  +
{\cal O }
(\delta^3) \: ,
\label{Gsimple}
\end{eqnarray}

\noindent
where $\psi (x)$ and $\psi'(x)$ are the psi-function and its first
derivative \cite{stegun}, respectively.

Substituting (\ref{Gsimple}) in (\ref{e8}),
we get the following expression for the
effective mass up
to second order in $\delta$, within the one-vertex two-point Green's function:

\begin{equation}
m_{\rm eff}^{2} = \mu^{2} + \frac{\lambda}{4 } 2 M^{2} - G_{(1v)}^{(2)}
\: ,
\label{mR1}
\end{equation}

\noindent
with $G_{(1v)}^{(2)}$ given by (\ref{Gsimple}).

The whole dependence of (\ref{mR1}) on the arbitrary mass parameter $M$
can be removed by requiring that \cite{stevenson}, \cite{rebhan}

\begin{equation}
\frac{\partial m_{\rm eff}^{2}}{\partial M^{2}} = 0 \: ,
\label{PMS}
\end{equation}

\noindent
at each order in the $\delta$-expansion. The condition (\ref{PMS})
fixes the value
of the mass parameter $M$ as being the one that leaves $m_{\rm eff}^{2}$
stationary (the PMS condition).

Using the variational procedure above, we get the following 
expressions for the mass
parameter $M$ at each order in $\delta$, for $\delta = 1$, and in
the high temperature limit ($I(m) \simeq \frac{T^{2}}{12}$, in
Eq. (\ref{Im2})):

\begin{equation}
2 M^{2} = \left\{
\begin{array}{ll}
\frac{T^{2}}{12} \exp \left[ 2 \psi(3) - \psi(1) \right] \: , & \mbox{ up to
order $\delta$} \\
\frac{T^{2}}{12} \exp \left[ 2 \psi(3) - \psi(1) - \sqrt{\psi'(1) - 4 \psi'(
3)} \: \: \right] \: , & \mbox{up to order $\delta^{2}$}
\end{array}
\right. \: ,
\label{2M2}
\end{equation}

Using (\ref{2M2}) for $M^{2}$ back in (\ref{mR1}),
we get the following expression for
the effective mass at finite temperature, up to orders $\delta$
and $\delta^{2}$, respectively:

\begin{equation}
m_{\rm eff}^{2}(T) = \left\{
\begin{array}{l}
\mu^{2} + \frac{\lambda}{4 } \frac{T^{2}}{12} \exp \left[ 2 \psi(3) -
\psi(1) \right] \: ,  \\
\mu^{2} + \frac{\lambda}{4 } \frac{T^{2}}{12} \exp \left[ 2 \psi(3) -
\psi(1) - \sqrt{\psi'(1) - 4 \psi'(3)} \: \: \right]
\left(1 + \sqrt{\psi'(1) -
4 \psi'(3)} \:\: \right)  \: .
\end{array}
\right.
\label{mR2}
\end{equation}

It is easy to show that (\ref{mR2}) must converge to the usual
1-loop approximation
for the finite temperature effective mass \cite{dolan}. {}From
(\ref{G1}), we can write

\begin{equation}
m_{\rm eff}^{2} = \mu^{2} + \frac{\lambda}{4 } M^{2} \frac{ (2 \delta + 2) !}
{2^{\delta} \delta !} \left[ M^{-2} I(m) \right]^{\delta} \: ,
\label{mRusual}
\end{equation}

\noindent
such that, in the high temperature limit, for $I(m) \simeq \frac{T^{2}}{12}$
and for $\delta =1$, we get the usual result,
$m_{\rm eff}^{2}(T) \simeq \mu^{2} + \lambda \frac{T^{2}}{4}$. However it is
remarkable that even for $\delta = 1$ and at lowest order, the expansion
(\ref{mR1})
is still consistent with the usual result
obtained via loop expansion. From (\ref{mR2}),
at first order in $\delta$ (the first
term in the right hand side in (\ref{mR2})), using the numerical values for the
$\psi$-functions \cite{stegun}, we get $m_{\rm eff}^{2}(T) \simeq \mu^{2} +
\lambda ' \frac{T^{2}}{4}$, where $\lambda ' =
\frac{\lambda}{12} \exp[2 \psi(3) -\psi(1)]\simeq
0.94 \lambda$. At second order in $\delta$, we have: $m_{\rm eff}^2(T)
\simeq \mu^2 + \lambda'' \frac{T^2}{4}$, where, from (\ref{mR2}), $\lambda''
= \frac{\lambda}{12} \exp \left[ 2 \psi(3) -
\psi(1) - \sqrt{\psi'(1) - 4 \psi'(3)} \: \: \right]
\left(1 + \sqrt{\psi'(1) -
4 \psi'(3)} \:\: \right) \simeq 0.91 \lambda$. With the consistent 
evaluation of terms of higher order
in $\delta$ with the introduction of higher order-vertex terms, it is
expected that $\lambda ' (\lambda'') \to \lambda$.

It is interesting to see that, from (\ref{Im1}) and (\ref{Im2}),
the loop integrals are written
with propagators carrying the extra factor $\frac{\lambda}{4 }
2 M^{2}$. However the variational condition, Eq. (\ref{PMS}), 
used to fix the value
of $M$, makes it possible to express the propagators with a finite
temperature mass. At first order in $\delta$,
using (\ref{Gsimple}) in (\ref{mR1}) we
obtain that $2 M^{2} = I(m) \exp[2 \psi(3) -
\psi(1)]$, where $m^{2} = \mu^{2} + \frac{\lambda}{4 } 2 M^{2}$.
If one redefines the coupling $\lambda$ as $\lambda ' = \frac{\lambda}{12}
\exp[2 \psi(3) -\psi(1)] \simeq 0.94 \lambda$ (or $\lambda'' \simeq 0.91
\lambda$, at order $\delta^2$), we then get, from the first order in $\delta$
term of (\ref{Gsimple}) substituted in (\ref{mR1}) and for $\delta=1$,

\begin{equation}
m_{\rm eff}^{2} = \mu^{2} + 3 \lambda ' I(m_{\rm eff}) \: ,
\label{gap1}
\end{equation}

\noindent
where the $I(m_{\rm eff})$ term can be expanded as in (\ref{Im2}). 
We recognize Eq. (\ref{gap1}) as a gap equation. Eq. (\ref{gap1}) 
is similar to the gap equation
in the $\lambda \phi^{4}$ model, obtained by incorporating, in the loop
expansion, the largest infrared
divergences, summing up the so called daisy (or superdaisy) diagrams 
\cite{dolan,quiros}.

If one expands (\ref{gap1}) in the high temperature limit,
we can obtain an approximate
equation for $m_{\rm eff}$

\begin{equation}
m_{\rm eff}^{2} = \mu^{2} + \frac{\lambda '}{4} T^{2} - 
3 \frac{\lambda '}{4 \pi}
T m_{\rm eff} \:,
\label{gap2}
\end{equation}

\noindent
from which we obtain, assuming $m_{\rm eff} \ge 0$, the solution (valid up to
${\cal O} (\frac{\lambda^{2} T^{2}}{m_{\rm eff}^{2}})$ )

\begin{equation}
m_{\rm eff} = - 3 \frac{\lambda '}{8 \pi} T + \sqrt{\left(
\frac{3 \lambda ' T}{
8 \pi}\right)^{2} + \mu^{2} + \frac{\lambda '}{4} T^{2}}  \: ,
\label{gap3}
\end{equation}

\noindent
which is in accordance with the result obtained in \cite{quiros},
by considering the contribution of superdaisy diagrams in the gap equation,
taking into account the leading infrared contributions at high temperatures.

\section{Conclusions}

The use of the
$\delta$-expansion, as shown in \cite{rebhan}, for the
particular case of the massless scalar $\lambda (\phi^{4})_{4}$ theory
with $O(N)$ global symmetry, at finite temperature, reproduces quite
well the exact result ($N \rightarrow \infty$ limit) for the gap equation,
up to order $\delta^{2}$. These results confirm our conclusions in that they
show
that the use of the PMS condition in the $\delta$-expansion is
self-consistent and is able to lead to nonperturbative results.

It would also be interesting to use the $\delta$-expansion for evaluating
higher order corrections for effective potentials at finite temperature, in
connection with the program of resummation, which has been an important
problem in the recent literature (see, for example, \cite{quiros} and
references therein). Work in this direction is in progress.

The
$\delta$-expansion has also been employed in the evaluation of critical
exponents \cite{exponents}, using exactly its ability of exploring the
infrared region, for $T$ close to $T_c$, the critical temperature in
spontaneously broken theories. A variant of the $\delta$-expansion used
here, called the linear $\delta$-expansion \cite{okopi}, has proven to be a
powerful tool for studying vacuum contributions on self-energies and in
energy densities of very different field theories (for an example, see for
instance \cite{krein}). The version of the $\delta$-expansion used in this
paper, usually called the non-linear or logarithmic $\delta$-expansion,
shares many properties with the linear one, representing, therefore, a
promising method
for getting vacuum fluctuation contributions, not only quantum but also
thermal contributions, as we have briefly demonstrated in this paper.

\vspace{0.5cm}
\begin{center}
{\large \bf Acknowledgements}
\end{center}

\vspace{1.0cm}
This work was partially
supported by Conselho Nacional de
Desenvolvimento Cient\'{\i}fico e Tecnol\'ogico - CNPq (Brazil).

\end{document}